\documentclass[pra ,showpacs,showkeys,twocolumn]{revtex4-1}

\usepackage{graphicx}
\usepackage{latexsym}
\usepackage{amsmath}
\usepackage{amssymb}
\usepackage{amsfonts}
\usepackage{color}
\usepackage{textcomp}
\usepackage[normalem]{ulem}

\newcommand{\jj}{\mathbf{j}}
\newcommand{\xx}{\mathbf{x}}

\newcommand{\Meff}{M_\textrm{eff}}

\begin{document}


\title{Vortices in nonequilibrium photon condensates}

\author{Vladimir Gladilin}
\author{Michiel Wouters}

\affiliation{TQC, Universiteit Antwerpen, Universiteitsplein 1,
B-2610 Antwerpen, Belgium}

\date{\today}

\begin{abstract}
We present a theoretical study of vortices in arrays of photon
condensates. Even when interactions are negligible, as is the case
in current experiments, pumping and losses can lead to a finite
vortex core size. While some properties of photon
condensate vortices, such as their self-acceleration and the
generation of vortex pairs by a moving vortex, resemble those in
interacting polariton condensates far from equilibrium, in several
aspects they differ from previously studied systems: the
vortex core size is determined by the balance between pumping and
tunneling, the core appears oblate in the direction of its motion
and new vortex pairs can spontaneously nucleate in the core
region.
%
%
\end{abstract}

\maketitle

{\em Introduction --} The experimental study of planar optical
systems has spurred the interest in studying the properties of light
from the point of view of quantum fluids \cite{carusotto13}.
Superfluidity \cite{amo09}  and quantized vortices
\cite{lagoudakis08} as one of its hallmarks \cite{bec_book} have
developed into a rich research domain with recent studies addressing
the Berezinskii-Kosterlitz-Thouless phase transition
\cite{caputo2016,wachtel,sieberer} , the Kibble-Zurek mechanism
\cite{comaron18,verstraelen20,kulczykowski17} and
Kardar-Parisi-Zhang dynamics of the superfluid phase
\cite{squizzato18,gladilin14,he15,altman15,sieberer16} . Moreover,
planar optical systems have been identified as promising platforms
to perform analog computation, with potential applications in NP
hard optimization problems. These ideas have been implemented in
exciton-polariton condensates
\cite{berloff17,lagoudakis17,kalinin18}, parametric oscillators
\cite{mcmahon16,yamamoto17}, coupled laser arrays \cite{tradonsky19}
and a proposal was done for coupled Bose-Einstein condensates (BECs)
of photons in a dye filled cavity \cite{kassenberg20}.

This latter photon condensate system
\cite{klaers10,marelic16,nyman18,greveling18} is in good
approximation an ideal Bose gas, which at first sight limits its
possibilities for the study of superfluidity. The ideal Bose gas has
a zero Landau critical velocity and has no well-defined quantized
vortices because the core size becomes as large as the entire system
\cite{bec_book}. This is related to the infinite compressibility of
the ideal Bose gas, which leads through the fluctuation-dissipation
relation to large density fluctuations \cite{klawunn11} that prevent
the emergence  of a well defined phase degree of freedom. However,
because of cavity mirror losses, which are compensated by external
pumping, photon condensates do differ from the ideal Bose gas.
Thanks to the interplay between pumping and dissipation, density
fluctuations can be reduced so that the photon phase becomes well
defined \cite{gladilin20}.

Experimentally, it has become possible to create arrays of photon
condensates by controllably coupling them through
tunneling \cite{kassenberg20,dung17}. In such systems a well defined
phase can be combined with spatial variations, which opens up the
possibility to study quantized vortices. Previous studies of
vortices \cite{gladilin17,gladilin19} and the BKT transition in
nonequilibrium polariton condensates \cite{wachtel,sieberer} have
shown that deviations from nonequilibrium strongly affect the
properties of single vortices as well as their interactions and
annihilation rates. In these works, an intrinsic nonlinearity was
included, such that the driving and dissipation formed a correction
to the equilibrium vortex structure. In photon BECs, in contrast,
driving and dissipation dominate the vortex structure and the
deviation from equilibrium becomes essential in order to obtain a
finite vortex core size.

{\em Model --} One of the advantages of photon condensates is that
their microscopic physics is well understood. The thermalization of
the photons through repeated emission and absorptions by the dye
molecules is achieved thanks to the Kennard-Stepanov relation
\cite{kennard,stepanov,moroshkin14} that sets a correspondence
between the emission ($B_{21}$) and absorption ($B_{12}$)
coefficients: $B_{12} = e^{\beta \Delta} B_{21}$, where $\Delta$ is
the detuning between the cavity and the dye transition frequencies
and $\beta=1/(k_B T)$ is the inverse temperature. For a single mode
system, rate equations give a good description of the photon density
dynamics \cite{schmitt18,verstraelen19}. For the extension to coupled
cavities, the simplest model is given by a generalized
Gross-Pitaevskii equation (gGPE) for the photon amplitudes
\cite{gladilin20}:
\begin{align}
i \hbar \frac{\partial \psi(\xx)}{\partial t}  &=  - (1-i\kappa) J  \sum_{\xx' \in \mathcal N_\xx} \psi({\xx'})  + V(\xx) \psi(\xx) \nonumber \\
&+ \frac{i}{2}\left[ B_{21}M_2(\xx) - B_{12}
M_1(\xx)-\gamma\right]\psi(\xx) . \label{eq:gGP}
\end{align}
It describes photons with loss rate $\gamma$ subject to an external
potential $V$ that hop to the nearest neighbor cavities, labeled
by $\xx'$, at tunneling rate $J$. The photons are coupled to the
dye molecules whose ground (excited) state occupation is denoted by
$M_{1(2)}$ satisfying at all times $M_1(\jj)+M_2(\jj)=M$, where $M$
is the number of dye molecules at each lattice site. The
Kennard-Stepanov relation gives rise to energy relaxation with
dimensionless strength
\begin{equation}
\kappa= \frac{1}{2} \beta B_{12} M_1. \label{eq:kappa}
\end{equation}
Our model does not include photonic interaction, which is quite
negligible in recent experiments \cite{randonji18}, except for a
slow thermo-optical nonlinearity \cite{alaeian17}, which does not
affect the physics over the time scales relevant for this study.

Equation \eqref{eq:gGP} is deterministic and it fails to
describe the fluctuations of the photon gas. These are included by
adding a unit complex number with random phase to the field
amplitude at the rate of the spontaneous photon emissions
\cite{henry82,verstraelen19}.

 The evolution of the number of excited molecules due to
 interactions with the photons is opposite to the change in number
 of photons due to emission (both deterministic and stochastic),
 absorption and energy relaxation. In order to compensate for the
 loss of energy in the system, external excitation with a pumping
 laser is needed to balance the photon losses:
 $P_{\rm pump} =\gamma \bar n$, where $\bar n$ is the targeted mean
 photon number.

For a single photonic mode, two different regimes can be identified
\cite{klaers12,schmitt14} in the parameter space of mean photon
number $\bar n$ and effective molecular reservoir size $\Meff = (
M+\gamma e^{-\beta \Delta}/B_{21})/[2+2\cosh(\beta \Delta)]$: (i) a
large fluctuation `grandcanonical' regime for small photon number
$\bar n^2 \ll M_{\rm eff}$, where $\langle (n-\bar n)^2 \rangle
\approx \bar n^2 $ and (ii) a small fluctuation `canonical' regime
for $\bar n^2 \gg M_{\rm eff}$, were $\langle (n-\bar n)^2 \rangle
\approx  M_{\rm eff} \ll \bar n^2 $.
 Only in the latter case, a well defined phase of the condensate
emerges. In the other case, frequent fluctuations to zero density
lead to reinitialization of the condensate with an arbitrary phase.
In this regime, the phase is not a slow degree of freedom with its
own effective dynamics as in the case of superfluids. This is
consistent with the thermal equilibrium nature of the ideal Bose
gas, which does not show real superfluid behavior.

In the canonical regime, when two or more photon modes are coupled
by tunnelling, the density fluctuations are determined by a
competition between tunnelling and pumping/losses \cite{gladilin20}.
In the absence of losses, the particle exchange causes density
fluctuations to go to the large grandcanonical value. The addition
of pumping and losses instead tends to fix the instantaneous local
density to the targeted one and suppresses density fluctuations. The
phase can then still be a good long wavelength degree of freedom. It
is only in this parameter regime that well defined vortices can be
studied.

{\em Results -- } Numerical simulations were performed on a
relatively large lattice of $101 \times 101$ coupled photon
condensates. An external potential of magnitude $V=-J$ at the border
of the lattice, and $V=-2 J$ in the corners is applied in order to
make the condensate density homogeneous. In the absence of this
potential offset, the condensate density profile becomes highly
inhomogeneous, which encumbers the analysis.

\begin{figure} \centering
\includegraphics*[width=0.8\linewidth]{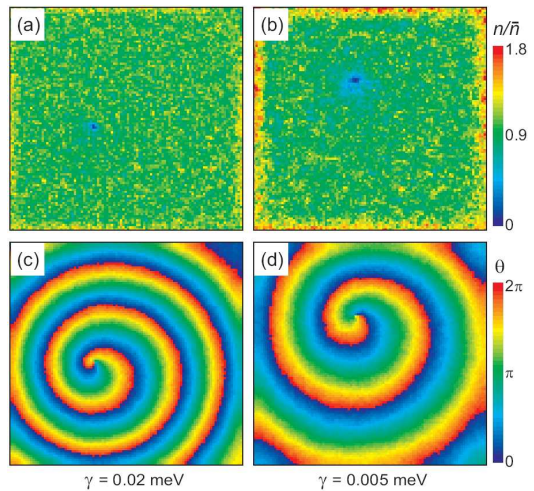}
\caption{ Snapshots of the photon density n [panels (a) and (b)] and
the phase of the order parameter $\theta$ [panels (c) and (d)] for
 $B_{21} = 10^{-6}$
meV ($\kappa$ = 0.763), $J$ = 0.02 meV and two different values of
the loss rate: $\gamma$ = 0.02 meV [panels (a) and (c)] and $\gamma
$ = 0.005 meV [panels (b) and (d)]. The other parameters of the
simulation are $k_B T=25$ meV, $\Delta/(k_B T)=-7.2$, $M=5\times
10^{10}$, $\bar n=5\times 10^{4}$.}  \label{f1}
\end{figure}

Two examples of vortices in an array of photon condensates are shown
in Fig.~\ref{f1}. The simulations were started with a phase winding
of $2\pi$ and then evolved until the density and phase patterns
around the vortex core were stabilized. In both simulations, it is
seen from the spiral form of phase pattern that in addition to the
azimuthal flow, there is a radial flow out of the vortex core. This
phenomenon is well known from studies of the complex Ginzburg-Landau
equation~\cite{aransonRMP02} and theory of vortices in
nonequilibrium polariton condensates \cite{gladilin17,gladilin19}.
The physical origin of the outward currents is the photon density
suppression at the vortex core, leading to less radiative losses
while pumping remains the same. The resulting excess energy is
carried away by the radial photon currents.

In the usual case of interacting bosons at thermal equilibrium, the
vortex core size is determined by the interaction energy, which sets
the healing length. This length scale tends to infinity in the limit
of vanishing interactions. This limit is recovered in the present
simulations when losses go to zero. However, driving and dissipation
provide an alternative mechanism to restrict
the vortex core size: within their finite life time, the photons can
only travel over a limited distance. This reasoning suggests an
inverse proportionality between the vortex core radius and the loss
rate. Analytical considerations based on the linearized equations
for density and phase fluctuations \cite{gladilin20} lead to the
estimate for the vortex core radius $r_v \propto \sqrt{J/\gamma}$.

The increase of the vortex core radius with increasing $J/\gamma$ is
confirmed by the numerical simulations in Fig. \ref{f1} [compare
panel (a) with larger losses than in panel (b)]. Since the radial
currents serve to carry away the excess excitation from the vortex
core region, they decrease with decreasing pumping and losses. This
is seen from the larger distance between the spiral arms in panel
(d) as compared to panel (c). The reduction of stimulated relaxation
of dye molecules in the vortex core region leads to a local increase
of the number of excited molecules (see supplementary information,
Fig. s1). In line with the increased core region for smaller losses,
the size of this molecular ``hot spot'' increases with decreasing
losses.

\begin{figure} \centering
\includegraphics*[width=0.8\linewidth]{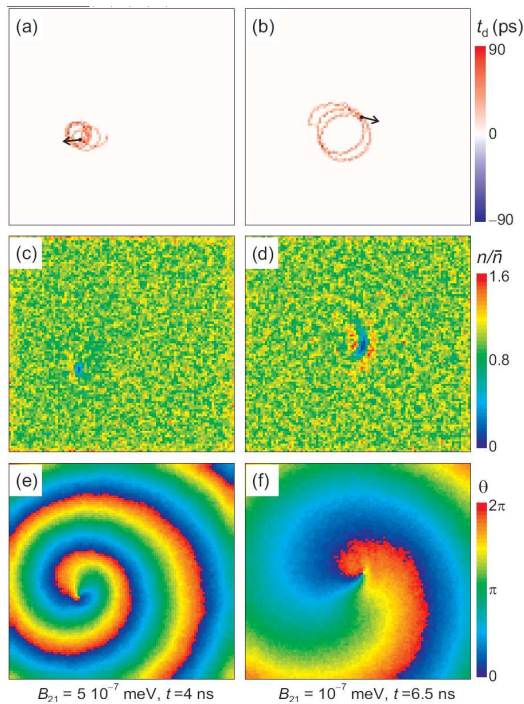}
\caption{(a-b) Vortex trajectories: the color codes the time the
core dwells in a given pixel with the sign of the dwelling time
representing the vorticity sign. The arrow indicates
the velocity at the final time of the simulation. Panels (c-d) show
the photon density profile at this instant, panels (e-f) the phase
profile. Parameters are the same as in Figs. \ref{f1}(a) and (c),
except for $B_{21}$. } \label{f2}
\end{figure}

So far, we have only discussed the density and phase profiles of the
vortex, but also the motion of the vortex core itself appears to be
interesting, as is illustrated in Fig. \ref{f2} (a) and (b). In
contrast to an equilibrium quantum fluid, where vortices move along
with the condensate, the vortex core now shows self-acceleration.
This phenomenon has been described previously in the context of the
cGLE \cite{aransonRMP02} and in a related phenomenological model for
polariton condensates \cite{gladilin17,gladilin19}. The present
simulations show that the acceleration of the vortex core also
appears in the absence of interparticle interactions.

The vortex follows to a good approximation a circular trajectory.
Deviations are caused by the fluctuations due to spontaneous
emission and by interactions with the sample boundaries. The
curvature of the trajectory can be understood from the Magnus effect
\cite{hall56,sonin97}  that is a consequence of a nonzero vortex
core velocity $v_{\rm vc}$ with respect to the surrounding
condensate (due to self-acceleration). Comparing panels (a) and (b)
of Fig. \ref{f2}, it is clear that the curvature of the trajectory
decreases with decreasing emission coefficient $B_{21}$. This
coefficient not only affects the conversion between photons and
excited molecules, but also the energy relaxation parameter
$\kappa$, cf. Eq. \eqref{eq:kappa}. We have verified in other
simulations (see supplementary information, Figs. s2 and s3) by
independently varying $B_{21}$ and $\kappa$ (could be physically
implemented by changing the temperature) that the increase in the
curvature of the trajectory is mainly due to the increase of
$B_{21}$ and not of $\kappa$. Some intuition for this dependence can
be obtained by considering the ratio $J/B_{21}$, which expresses the
tunneling strength in the energy scale of the interaction between
photons and molecules. Increasing $J/B_{21}$ can be thought of as an
increase in resolution within the continuum limit of the lattice
model. This change is then accompanied by a magnification of the
vortex trajectory picture. {When, on the other hand, $\kappa$ is
varied independently, the curvature of the vortex trajectory appears
to remain nearly unaltered.}

Apart from the change in trajectory curvature, reducing $B_{12}$
leads to an increase in size and a deformation of the core. In panel
(d) the core is significantly elongated in the direction
approximately perpendicular to the direction of the vortex motion,
that is indicated by the arrow in panel (b) (see also Fig. s4 in
Supplementary information). This contrasts with vortices in
superconductors, where due to a long healing time of the order
parameter the core of a fast moving vortex elongates in the
direction of its motion~\cite{vodolazov2007,vdvondel2011}. Apart
from the core deformation, also photon density maxima behind
and in front of the core appear. In these high density regions, the
photon losses are higher, such that the radial currents emitted by
the vortex core are partially drained here. This is reflected in the
phase profile [compare panel (f) to panel (e)] as a slower winding
of the spiral arms.

\begin{figure} \centering
\includegraphics*[width=0.8\linewidth]{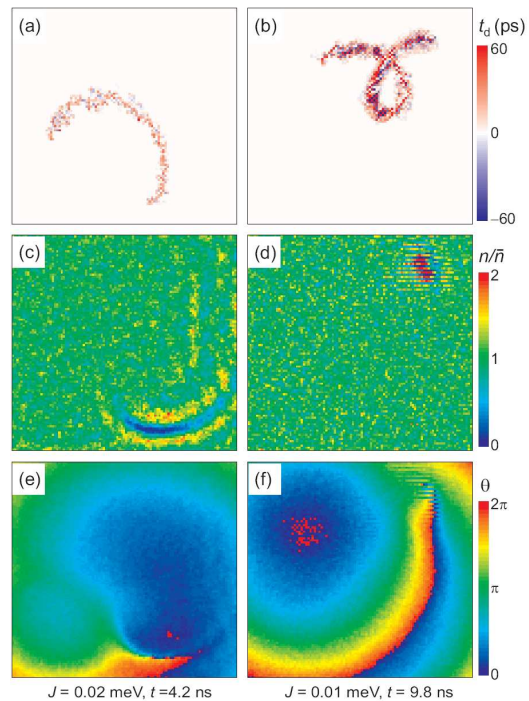}
\caption{Formation of baby vortices inside the core when the
tunneling rate $J$ is reduced. The panels give the same information
as in Fig. \ref{f2}. Parameters are the same as in Fig. \ref{f1} (a)
and (c), except for a reduced value of $J$ for panels (b), (d) and
(f) and a significantly smaller $B_{21}=2.5\times 10^{-8}$ meV for
all panels.} \label{f3}
\end{figure}

More exotic physics appears when the emission coefficient $B_{21}$
and, correspondingly, the energy relaxation parameter $\kappa$ are
further reduced. The vortex core then turns out to be unstable as
most clearly illustrated in the right hand side panels (b,d,f) of
Fig. \ref{f3}. In panels (a,b) an area with both negative and
positive vorticity can be recognized, while the
imprinted vorticity is positive, which is reflected in the long
distance behavior of the phase. At short distances on the other
hand, one sees the formation of baby vortex-antivortex pairs.
Correspondingly in the density profile, see panel (d), the vortex
core is no longer a simple minimum of the density, but it shows
large density variations. Because the vortex core is now filled, the
radial currents become much weaker, such that the spiral form of the
vortex almost disappears.

The reason for the formation of baby vortices is that the
low density in the vortex core enhances the phase fluctuations
caused by spontaneous emission. Loosely speaking, one could
compare these baby vortices with the spontaneous formation of
vortices in the Berezinskii-Kosterlitz-Thouless (BKT) phase
transition, which occurs when the density drops below a (temperature
dependent) critical value \cite{bec_book}.

\begin{figure} \centering
\includegraphics*[width=0.8\linewidth]{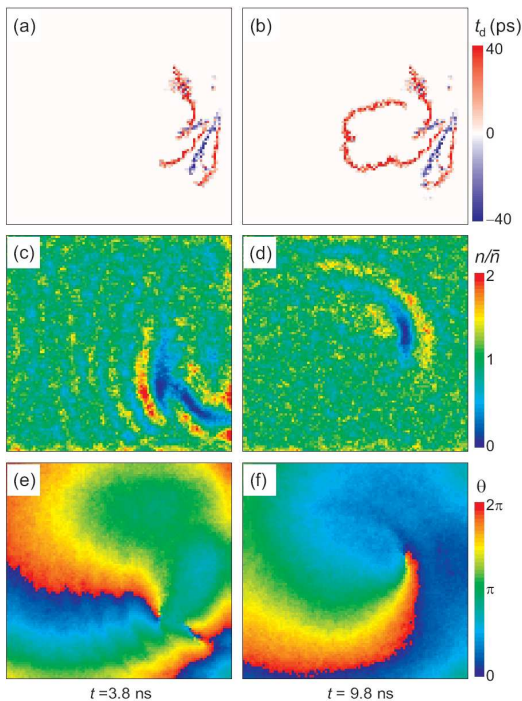}
\caption{ Trajectories of vortices (red) and antivortices (blue) up
to 3.8 ns (a) and 9.8 ns (b), when the original vortex approaches
the boundary . Formation of new pairs at the boundary and their
recombination can be seen. The density and phase after 3.8 ns are
shown in panels (c,e), where in addition to the original vortex, a
clear vortex-antivortex pair is visible. Panels (d,f) show density
and phase after 9.8 ns, when only a single vortex survives after it
has moved away from the edge of the cavity array. Parameters are the
same as in Fig. \ref{f1}(b) and (d), except for $B_{21}=2.5\times
10^{-8}$ meV.} \label{f4}
\end{figure}

In Fig. \ref{f4}, we show a situation where a vortex approaches the
boundary of the sample. The appearance of blue pixels in panels (a)
and (b) demonstrates that new vortex-antivortex pairs are generated.
Density and phase profiles at the time when the vortex is close to
the boundary are shown in panels (c,e). A new vortex-antivortex pair
is clearly visible. Panel (b) shows that the original vortex
actually recombines with one of the newly created antivortices and
that the other member of the pair continues as the only remaining
vortex when it moves away from the boundary, see panels (d,e). The
production of vortex-antivortex pairs at the boundary resembles the
creation of baby vortices inside the vortex core, but it already
happens at larger ratio $J/\gamma$ (where the vortex core is more
stable), with the boundary facilitating the formation of new pairs.

The generation
of vortex-antivortex pairs by fast moving vortices was already seen
in simulations for polariton condensates
\cite{gladilin19,prb19},
where we used a continuum model that was discretized only for the
numerical solution of the partial differential equation. In
contrast, for photon condensates, it is crucial to use a discrete
set of coupled cavities in order to keep the density fluctuations in
check and to limit the vortex core size.

{\em Conclusions and outlook --} Our study of arrays of coupled
photon condensates has shown that the formation of quantized
vortices with a well defined core is possible in the canonical
regime if the tunneling strength is small enough with respect to the
loss rate. Because the vortex properties are entirely determined {by
the interplay between tunneling and pumping},  they feature unusual
properties when compared to vortices in traditional superfluids.

We have recovered several features that were previously obtained for
nonequilibrium polariton condensates, such as radially diverging
particle flows from the vortex core \cite{aransonRMP02},
self-accelerated vortex motion \cite{gladilin17}
and the production of vortex-antivortex
pairs in interactions with the sample boundary \cite{gladilin19}.
Novel features are the vortex core instability with the formation of
``baby'' vortices and the elongation of the core that is
quasi-perpendicular to its velocity.

In analogy with polariton condensates \cite{gladilin17,gladilin19},
it is expected that the radial flows will lead to vortex-antivortex
repulsion and inhibit vortex-antivortex recombination. A preliminary
case study (see supplementary material Fig. s5) confirms this
expectation, but a comprehensive analysis of the
interactions between vortices is beyond the scope of the present
work.

 Our results pave the way for studies of photon condensates with
multiple vortices, which may be generated by phase imprinting, in
quenches through the Kibble-Zurek mechanism or thermally activated
as in the BKT transition. Finally, vortices may also play a role in
the formation of metastable states in proposed schemes for analog
solving of optimization problems \cite{kassenberg20}.

Our theoretical predictions rely upon a semiclassical model which is
based on microscopic physics that is well understood and should be
accessible to experiments where photon condensates can be coupled in
a controlled manner \cite{dung17,kassenberg20}.

{\em Acknowledgements --} We are grateful to Jan Klaers, Fahri
Ozturk, Martin Weitz and Wouter Verstraelen for stimulating
discussions. VG was financially supported by the grant
UA-BOF-FFB150168.

\newpage

\section*{Supplementary figures}
\newpage

\setcounter{figure}{0}    
\renewcommand\thefigure{s\arabic{figure}}

\begin{figure} \centering
\includegraphics[width=0.9\linewidth]{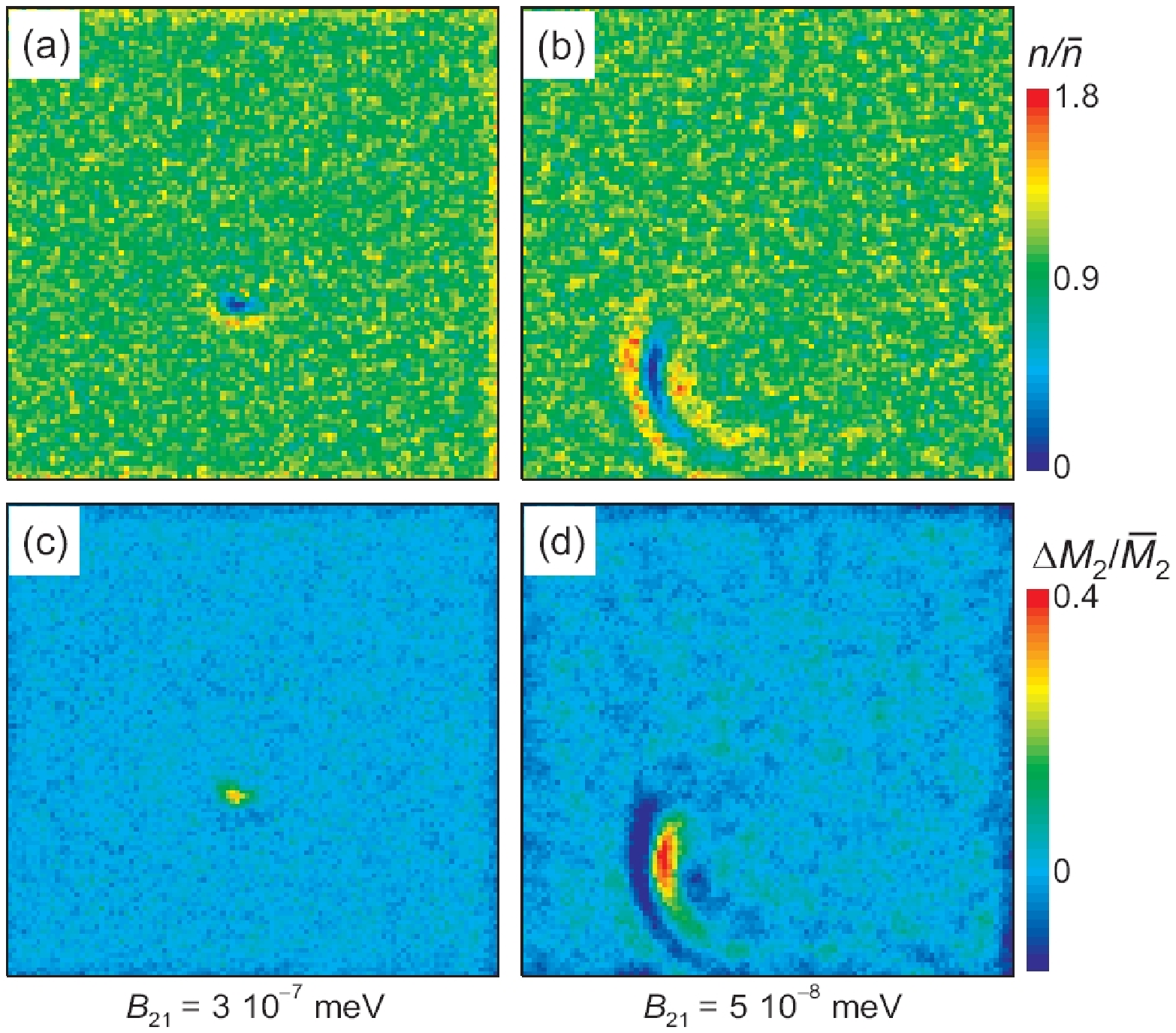}
 \caption{Snapshots of the photon density $n$
[panels (a) and (b)] and the distribution of deviations of the
number of excited molecules from its average value $\Delta M_2\equiv
M_2-\bar M_2$ [panels (c) and (d)] for $k_B T=25$ meV, $\Delta/(k_B
T)=-7.2$, $M=5\times 10^{10}$, $\bar n=5\times 10^{4}$, $J=0.02$
meV, $\gamma=0.02$ meV and two different values of the emission
coefficient: $B_{21}= 3\times 10^{-7}$ meV ($\kappa=0.229$), [panels
(a) and (c)] and $B_{21}= 5\times 10^{-8}$ meV ($\kappa=0.038$)
[panels (b) and (d)]. There is a clear anticorrelation between the
distributions of $n$, on the one hand, and $M_2$, on the other
hand.} \label{sf1}
\end{figure}

\begin{figure} \centering
\includegraphics[width=0.9 \linewidth]{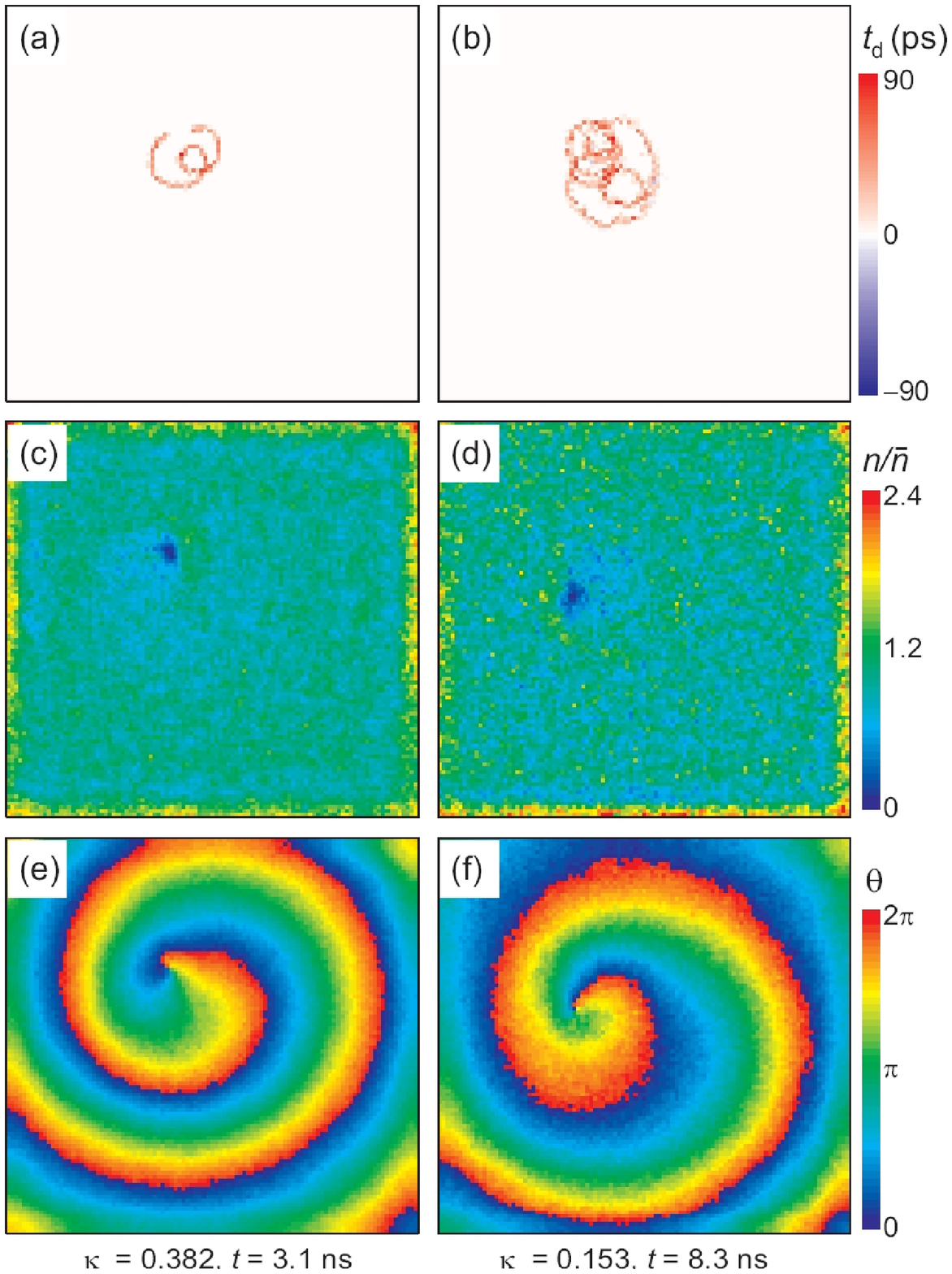}
\caption{Vortex trajectories within the time interval $(0,t)$
[panels (a) and (b)] and snapshots of the photon density $n$ [panels
(c) and (d)] and the phase of the order parameter $\theta$ [panels
(e) and (f)] at the time moment $t$ for $\Delta/(k_B T)=-7.2$,
$M=5\times 10^{10}$, $\bar n=5\times 10^{4}$, $J=0.02$ meV,
$\gamma=0.005$ meV, $B_{21}=5\times 10^{-7}$ meV and two different
values of $\kappa$, which is considered here as an independent
parameter: $\kappa=0.382$ [panels (a), (c) and (e)] and
$\kappa=0.153$ [panels (b), (d) and (f)]. A decrease of $\kappa$
results in an increase the of density and phase fluctuations,
induced by the Josephson junctions in the array of microcavities.
The size of the vortex core as well as the phase gradients,
corresponding to the radial superflows, and the shape of the vortex
trajectories are not significantly influenced by moderate variations
of $\kappa$.} \label{sf2}
\end{figure}

\begin{figure} \centering
\includegraphics[width=0.9 \linewidth]{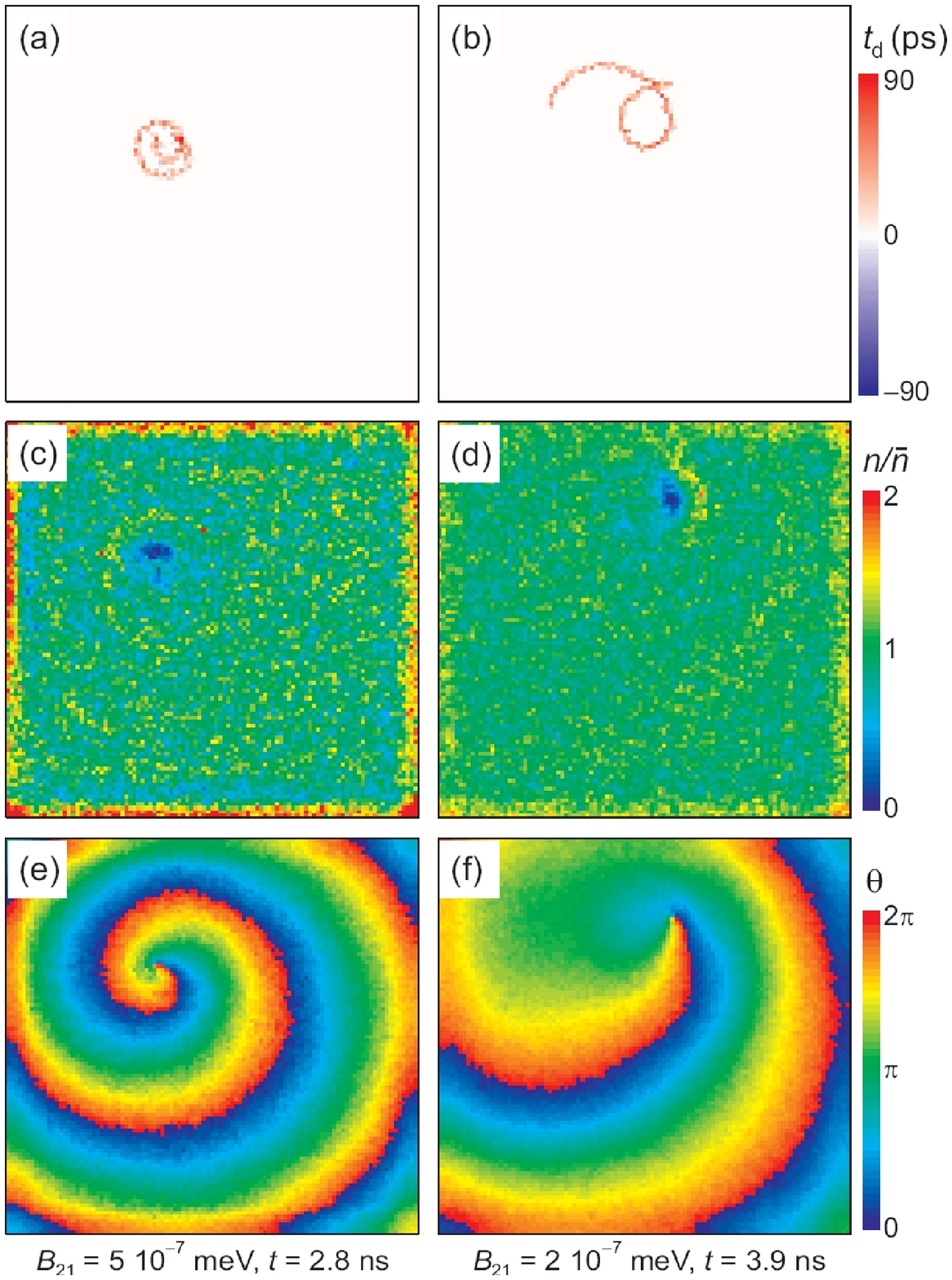}
\caption{Vortex trajectories within the time interval $(0,t)$
[panels (a) and (b)] and snapshots of the photon density $n$ [panels
(c) and (d)] and the phase of the order parameter $\theta$ [panels
(e) and (f)] at the time moment $t$ for $\Delta/(k_B T)=-7.2$,
$M=5\times 10^{10}$, $\bar n=5\times 10^{4}$, $J=0.02$ meV,
$\gamma=0.005$ meV, $\kappa=0.153$ meV ($\kappa$ is considered here
as an independent parameter) and two different values of $B_{21}$:
$B_{21}=5\times 10^{-7}$ meV [panels (a), (c) and (e)] and
$B_{21}=2\times 10^{-7}$ meV [panels (b), (d) and (f)]. When
decreasing $B_{21}$, the exchange between the photon subsystem
inside each microcavity and the molecule reservoir slows down as
compared to the intercavity exchange due to the Josephson coupling.
This is accompanied by a reduction of phase gradients in the radial
direction and a decrease of the curvature of the vortex trajectory.
At the same time, a decrease of $B_{21}$ leads to the corresponding
weakening of the phase and density fluctuations, caused by the
spontaneous emission of photons in each microcavity.} \label{sf3}
\end{figure}

\begin{figure} \centering
\includegraphics[width=0.9 \linewidth]{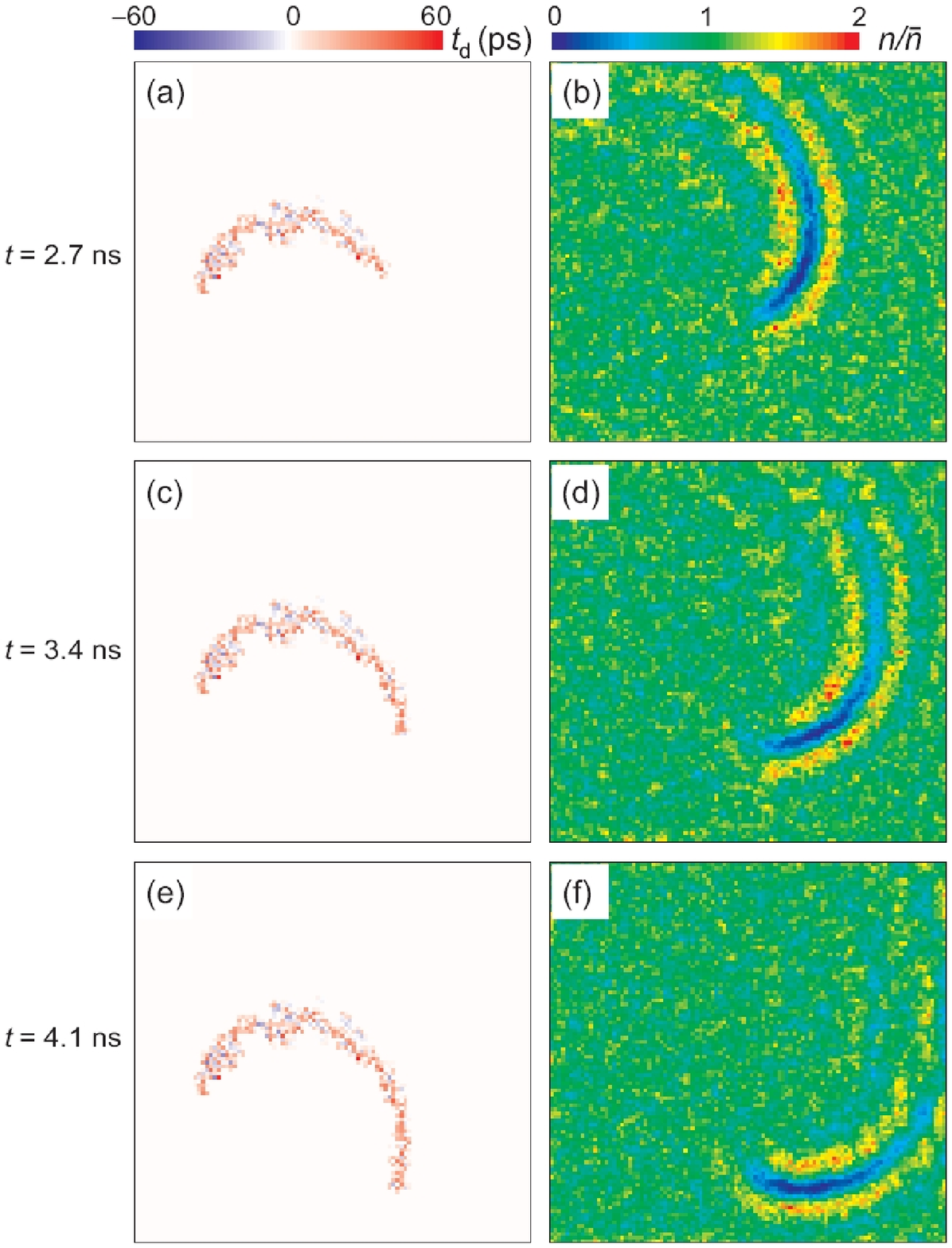}
\caption{Vortex trajectories within the time interval $(0,t)$
[panels (a), (c) and (e)] and snapshots of the photon density $n$
[panels (b), (d) and (f)] at the time moment $t$ for $k_B T=25$ meV,
$\Delta/(k_B T)=-7.2$, $M=5\times 10^{10}$, $\bar n=5\times 10^{4}$,
$J=0.02$ meV, $\gamma=0.02$ meV, $B_{21}=2.5\times 10^{-8}$ meV and
different $t$: $t=2.7$ ns [panels (a) and (b)], $t=3.4$ ns [panels
(c) and (d)], and $t=4.1$ ns [panels (e) and (f)]. The reorientation
of the elongated vortex core with time is seen to be retarded with
respect to the changes of the vortex motion direction.} \label{sf4}
\end{figure}

\begin{figure} \centering
\includegraphics[width=0.9\linewidth]{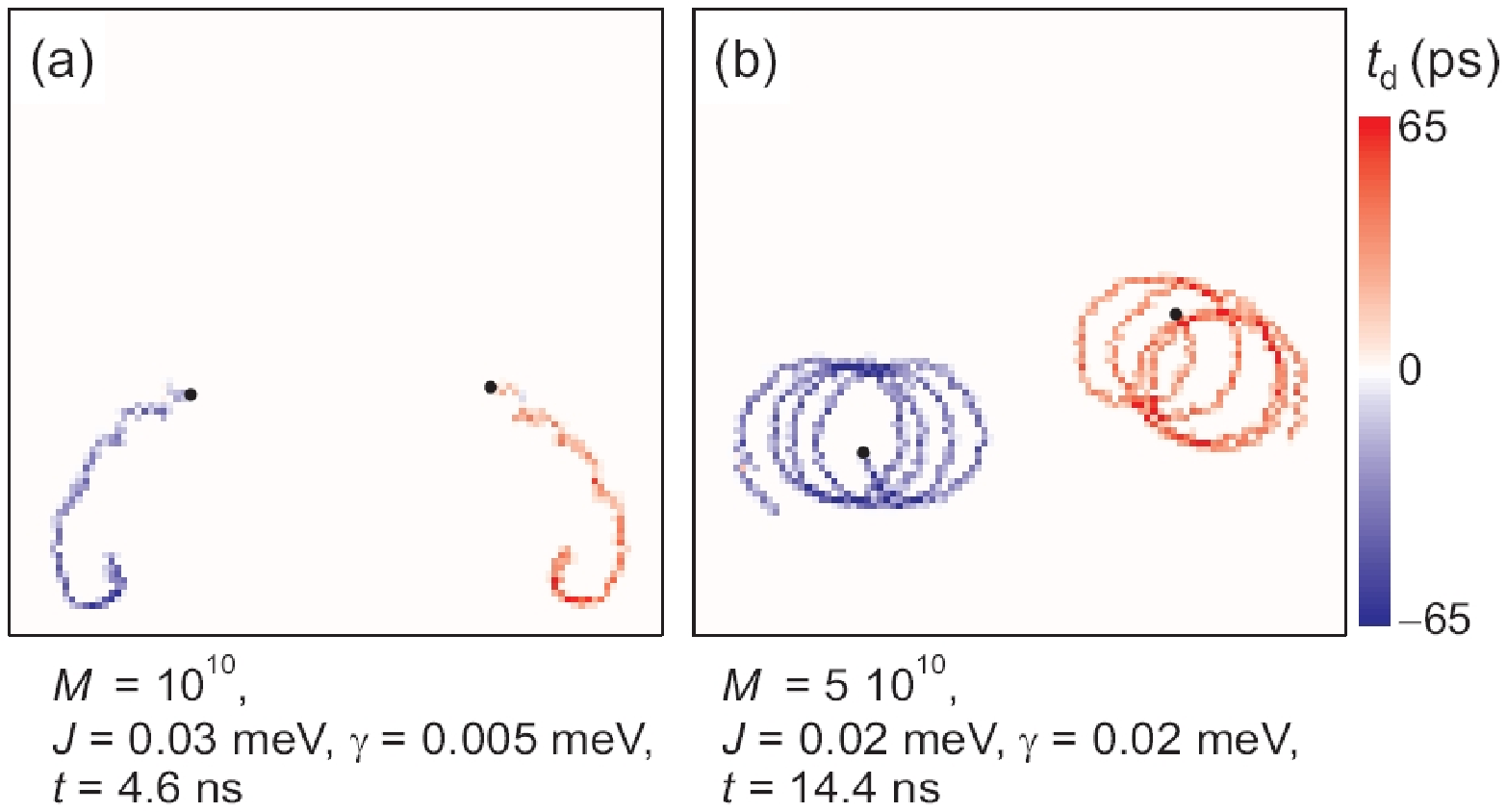}
 \caption{Vortex (red) and antivortex (blue) trajectories within the time interval $(0,t)$
 for $M=10^{10}$, $J=0.03$ meV, $\gamma=0.05$ meV, $t=4.6$
 ns (a) and $M=5\times 10^{10}$, $J=0.02$ meV, $\gamma=0.02$ meV,
 $t=14.4$ (b). Other parameters are $k_B T=25$ meV, $\Delta/(k_B
T)=-7.2$, $\bar n=5\times 10^{4}$, $B_{21}= 10^{-7}$ meV. The black
circles indicate the (anti)vortex positions at $t=0$. The overall
direction of vortex motion corresponds to the presence of a
(long-range) vortex-antivortex repulsion.} \label{sf5}
\end{figure}


\begin{references}

\bibitem{carusotto13} I. Carusotto and C. Ciuti, Reviews of Modern Physics {\bf 85}, 299 (2013).

\bibitem{amo09} A. Amo, J. Lefr\`ere, S. Pigeon, C. Adrados, C. Ciuti, I. Carusotto, R. Houdr\'e, E. Giacobino and A. Bramati, Nat. Phys. {\bf 5}, 805 (2009).

\bibitem{lagoudakis08} K. G. Lagoudakis, M. Wouters, M. Richard, A. Baas, I. Carusotto, R. Andr\'e, Le Si Dang and B. Deveaud-Pl\'edran, Nat. Phys. {\bf 4}, 706 (2008).

\bibitem{bec_book}  L. P. Pitaevski and S. Stringari, Bose-Einstein  condensation (Oxford University Press, 2016).

\bibitem{caputo2016}
D.~Caputo, D.~Ballarini, G.~Dagvadorj, C.~S. Mu{\~n}oz,
M.~De~Giorgi,L.~Dominici, K.~West, L.~N. Pfeiffer, G.~Gigli, F.~P.
Laussy, M. H. Szyma\'nska and D. Sanvitto, Nat. Mat. {\bf 17},145
(2018).

\bibitem{wachtel}  G. Wachtel, L. M. Sieberer, S. Diehl and E. Altman, Phys. Rev. B {\bf 94}, 104520 (2016).

\bibitem{sieberer} L. M. Sieberer, G. Wachtel, E. Altman and S. Diehl, Phys. Rev. B {\bf 94}, 104521 (2016).

\bibitem{comaron18} P.  Comaron,  G.  Dagvadorj,  A.  Zamora,  I.  Carusotto, N. P. Proukakis, and M. H. Szyma\'nska, Dynamical critical  exponents  in  driven-dissipative  quantum  systems, Phys. Rev. Lett. {\bf 121}, 095302 (2018).

\bibitem{kulczykowski17} M.  Kulczykowski  and  M.  Matuszewski, Phys. Rev. B {\bf 95}, 075306 (2017).

\bibitem{verstraelen20} M. Wouters and W. Verstraelen, Phys. Rev. A {\bf 101}, 043826 (2020).

\bibitem{squizzato18} D. Squizzato, L. Canet and A. Minguzzi, Phys. Rev. B \textbf{97}, 195453 (2018).

\bibitem{gladilin14} V. N. Gladilin, K. Ji, and M. Wouters, Phys. Rev. A \textbf{90}, 023615 (2014).

\bibitem{he15} L. He, L. M. Sieberer, E. Altman, and S. Diehl, Phys.
Rev. B 92, 155307 (2015).

\bibitem{altman15} E. Altman, L. M. Sieberer, L. Chen, S. Diehl, and J.
Toner, Phys. Rev. X 5, 011017 (2015).

\bibitem{sieberer16} L. M. Sieberer, M. Buchhold, and S. Diehl, Reports on
Progress in Physics 79, 096001 (2016).

\bibitem{berloff17} N. G. Berloff, M. Silva, K. Kalinin, A. Askitopoulos, J. D. T\"opfer, P. Cilibrizzi, W. Langbein and P. G. Lagoudakis, Nat. Mater. {\bf 16}, 1120 (2017).

\bibitem{lagoudakis17} P. Lagoudakis and N. G. Berloff,  New J. Phys. {\bf 19}, 125008 (2017).

\bibitem{kalinin18} K. Kalinin, P. G. Lagoudakis, N. G. Berloff, Phys. Rev. B  {\bf 97}, 094512 (2018).

\bibitem{mcmahon16} P. L. McMahon, A. Marandi, Y. Haribara, R. H, C. Langrock, S. Tamate, T. Inagaki, H. Takesue, S. Utsunomiya, K. Aihara, R. L. Byer, M. M. Fejer and H. Mabuchi, Y. Yamamoto,
Science {\bf 354}614 (2016).

\bibitem{yamamoto17} Y. Yamamoto, K. Aihara, T. Leleu, K. Kawarabayashi, S. Kako, M. Fejer, K. Inoue and H. Takesue NPJ Quant. Inf. {\bf 3}, 49 (2017).

\bibitem{tradonsky19} C. Tradonsky, I. Gershenzon, V. Pal, R. Chriki, A. A. Friesem, O. Raz and N. Davidson Science Advances. {\bf 5}, 4530 (2019).

\bibitem{kassenberg20} B. Kassenberg, M. Vretenar, S. Bissesar, J. Klaers, arXiv:2001.09828.

\bibitem{klaers10} J. Klaers, J. Schmitt, F. Vewinger,  and M. Weitz, Nature {\bf 468}, 545 (2010).

\bibitem{marelic16}  J.  Marelic,  L.  F.  Zajiczek,  H.  J.  Hesten,  K.  H.  Leung,E. Y. X. Ong, F. Mintert  and R. A. Nyman, N. J. Phys. {\bf 18}, 103012 (2016).

\bibitem{nyman18} B. T. Walker, L. C. Flatten, H. J. Hesten, F. Mintert, D. Hunger, A. A. P. Trichet, J. M. Smith and R. A. Nyman, Nat. Phys. {\bf 14}, 1173 (2018).

\bibitem{greveling18}  S. Greveling,  K. L. Perrier and D. van Oosten, Phys.Rev. A {\bf 98}, 013810 (2018).

\bibitem{klawunn11} M. Klawunn, A. Recati, L. P. Pitaevskii and S. Stringari, Phys. Rev. A {\bf 84}, 033612 (2011).

\bibitem{gladilin20} V. N. Gladilin and M. Wouters,
Phys. Rev. A {\bf 101}, 043814 (2020).


\bibitem{dung17} D. Dung, C. Kurtscheid, T. Damm, J. Schmitt, F. Vewinger, M. Weitz
and J. Klaers, Nat. Phot. {\bf 11}, 565 (2017).


\bibitem{gladilin17} V. N. Gladilin and M. Wouters, New J. Phys. {\bf 19}, 105005
(2017).

\bibitem{gladilin19} V. N. Gladilin and M. Wouters, J. Phys. A {\bf 52}, 1751 (2019).

\bibitem{kennard} E. H. Kennard, Phys. Rev. {\bf 11}, 29 (1918).

\bibitem{stepanov} B. I. Stepanov, Dokl. Akad. Nauk SSSR {\bf 112}, 839 (1957).
[Sov. Phys. Dokl. 2, 81 (1957)].

\bibitem{moroshkin14} P. Moroshkin, L. Weller, A. Sa\ss, J. Klaers and M. Weitz, Phys. Rev. Lett. {\bf 113}, 063002 (2014).

\bibitem{schmitt18} J.  Schmitt,  Journal  of  Physics  B, {\bf 51}, 173001 (2018).

\bibitem{verstraelen19} W. Verstraelen and M. Wouters, Phys. Rev. A
{\bf 100}, 013804 (2019).

\bibitem{randonji18} M. Radonji, W. Kopylov, A. Bala,  and A.
Pelster, N. J. Phys. {\bf 20}, 055014 (2018).

\bibitem{alaeian17} H.
Alaeian,  M.  Schedensack,  C.  Bartels,  D.  Peterseim and M.
Weitz, N. J. Phys. {\bf 19}, 115009 (2017).

\bibitem{henry82}  C. Henry, IEEE Journal of Quantum Electronics {\bf 18},
259(1982).

\bibitem{klaers12}  J.  Klaers,  J.  Schmitt,  T.  Damm,  F.  Vewinger,and M. Weitz,
Phys. Rev. Lett. {\bf 108}, 160403 (2012).

\bibitem{schmitt14}  J. Schmitt, T. Damm, D. Dung, F. Vewinger, J. Klaers and M. Weitz,
Phys. Rev. Lett. {\bf 112}, 030401 (2014).

\bibitem{aransonRMP02} I. S. Aranson and L. Kramer, Rev. Mod. Phys. {\bf 74}, 99
(2002).

\bibitem{hall56} H.E. Hall and W.F. Vinen, Proc. Roy. Soc. A {\bf 238}, 204 (1956).

\bibitem{sonin97} E. B. Sonin Phys. Rev. B {\bf 55}, 485 (1997).

\bibitem{vodolazov2007} D. Y. Vodolazov and F. M. Peeters, Phys. Rev. B 76,
014521 (2007).

\bibitem{vdvondel2011} J. Van de Vondel, V. N. Gladilin, A. V. Silhanek, W. Gillijns,
J. Tempere, J. T. Devreese, and V. V. Moshchalkov, Phys. Rev. Lett.
{\bf 106}, 137003 (2011).

\bibitem{prb19} V. N. Gladilin and M. Wouters, Phys. Rev. B {\bf 100}, 214506
(2019).

\end{references}
\end{document}